\documentclass[12pt]{article}
\usepackage{graphicx}% Include figure files
\usepackage{dcolumn}% Align table columns on decimal point
\usepackage{bm}% bold math
\usepackage{amssymb}   % for math
\usepackage{psfig}   % for math

\newcommand{\bvt}{\begin{verbatim}}
\newcommand{\evt}{\end{verbatim}}

\begin{document}

\title{\bf fastNLO: Fast pQCD Calculations \\ for PDF Fits}
\author{T. Kluge$^1$, K. Rabbertz$^2$, M. Wobisch$^3$\\ \, \\
\em     $^1$ DESY, Hamburg, Germany
  \\\em$^2$ University of Karlsruhe, Karlsruhe, Germany
  \\\em$^3$ Fermi National Accelerator Laboratory, Batavia, IL, USA}

\maketitle

\begin{abstract}
We present a method for very fast repeated computations of higher-order
cross sections in hadron-induced processes
for arbitrary parton density functions.
A full implementation of the method
for computations of jet cross sections
in Deep-Inelastic Scattering and in Hadron-Hadron Collisions
is offered by the ``fastNLO'' project.  
A web-interface for online calculations and user code
can be found at
{\tt http://hepforge.cedar.ac.uk/fastnlo/}.
\end{abstract}

\vskip-13.8cm
\hfill {hep-ph/0609285} \\
\phantom{a} \hfill{DESY 06-186}  \\
\phantom{a} \hfill{FERMILAB-CONF-06-352-E} \\

\vskip18cm

\noindent
To appear in the
{\sl Final Report of the QCD working group of the Tevatron-for-LHC Workshop}, 
eds. M. Carena and S. Mrenna (2006)
\vskip2mm
\noindent
A slightly shorter version will appear in
{\sl Proceedings of the XIV Workshop on Deep Inelastic Scattering, 
April 20-24, 2006 in Tsukuba, Japan}.

\newpage
% --------------------------------------------------------------
% --------------------------------------------------------------
% --------------------------------------------------------------
\section{Introduction}

The aim of the "fastNLO" project is to make the inclusion of jet data
into global fits of parton density functions (PDFs) feasible. 
Due to the prohibitive computing time required for the jet cross sections 
using standard calculation techniques,
jet data have either been omitted in these fits completely 
or they were included using a simple approximation.
The fastNLO project implements a method that offers exact and 
very fast pQCD calculations
for a large number of jet data sets 
allowing to take full advantage of their direct sensitivity 
to the gluon density in the proton in future PDF fits.
This includes Tevatron jet data beyond
the inclusive jet cross section and also
HERA jet data which have %successfully  
been used to determine the proton's gluon
density~\cite{Adloff:2000tq,Chekanov:2001bw,Chekanov:2002be,Chekanov:2005nn},
but which are ignored in current 
PDF fits~\cite{Alekhin:2005gq,Martin:2004ir,Pumplin:2002vw}.

% --------------------------------------------------------------
% --------------------------------------------------------------
% --------------------------------------------------------------
\section{Concept}

\subsection{Cross Sections in Perturbative QCD}

Perturbative QCD predictions for observables in 
hadron-induced processes depend on the strong coupling 
constant $\alpha_s$ and on the PDFs of the hadron(s).
Any cross section in hadron-hadron collisions 
can be written as the convolution of 
the strong coupling constant  $\alpha_s$ in order $n$,
the perturbative coefficient $c_{n,i}$ for the partonic
subprocess $i$,
and the corresponding linear combination of PDFs 
from the two hadrons $F_i$
which is a function of the  fractional hadron momenta
$x_{a,b}$ carried by the partons
\begin{equation}
\sigma(\mu_r,\mu_f) = \sum_{n,i}  \, c_{n,i}(x_a, x_b, \mu_r,\mu_f) 
\otimes 
\left[ \alpha_s^n(\mu_r) \cdot F_i(x_a,x_b,\mu_f) \right] \,.
\label{eq:fnmain}
\end{equation}
The PDFs and $\alpha_s$ also depend on the factorization and the 
renormalization scales $\mu_{f,r}$, respectively,
as does the perturbative prediction for the cross section
in finite order $n$.
An iterative PDF fitting procedure
using exact NLO calculations for jet data, 
based on Monte-Carlo integrations of~(\ref{eq:fnmain}), 
is too time-consuming.
Only an approximation of~(\ref{eq:fnmain}) is, therefore,
currently being used in global PDF fits.

% --------------------------------------------------------------
% --------------------------------------------------------------
% --------------------------------------------------------------
\subsection{A Simple Approach}

\begin{figure}[t]
\centerline{
   \psfig{figure=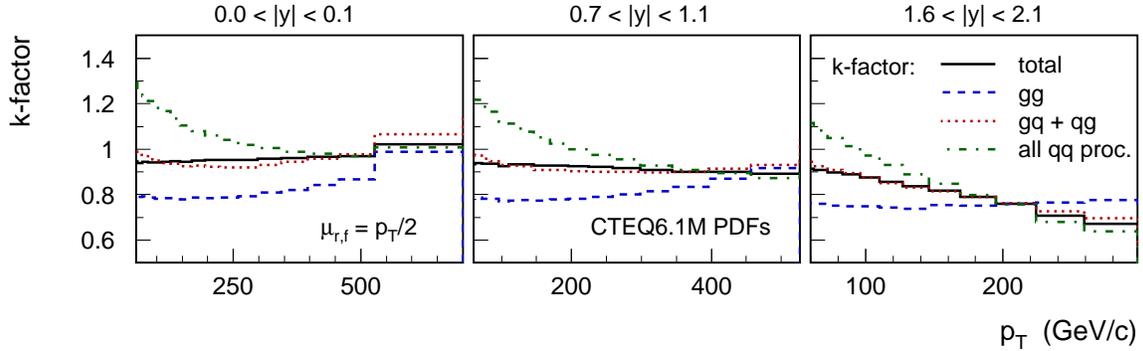,height=4.9cm}
}
   \caption{The $k$-factor for the inclusive $p\bar{p}$ jet cross section 
   at $\sqrt{s}=1.96$\,TeV as a function of $p_T$ at different rapidities $y$
   for the total cross section (solid line) and for different 
   partonic subprocesses:
   gluon-gluon (dashed), gluon-quark (dotted) and the sum of all
   quark and/or anti-quark induced subprocesses (dashed-dotted).
\label{fig:kfactor}}
\end{figure}

The ``$k$-factor approximation''
as used in~\cite{Martin:2004ir,Pumplin:2002vw}
parameterizes higher-order corrections
for each bin of the observable by a factor
$\displaystyle k = \frac{\sigma_{\rm NLO}}{\sigma_{\rm LO}}
= \frac{\sigma_{(2)}+\sigma_{(3)}}{\sigma_{(2)}}$
computed from the contributions 
with $n=2$ ($\sigma_{(2)}$) and $n=3$ ($\sigma_{(3)}$)
for a fixed PDF, averaged over all subprocesses~$i$.
In the iterative fitting procedure
only the LO cross section is computed
and multiplied with $k$ to obtain an estimate of 
the NLO cross section.
This procedure does not take into account that 
different partonic subprocesses can have largely 
different higher-order corrections.
Fig.~\ref{fig:kfactor} shows that the $k$-factors
for quark-only and gluon-only induced subprocesses
can differ by more than $\pm20\%$ from the average.
The $\chi^2$ is therefore minimized under an incorrect assumption
of the true PDF dependence of the cross section.
Further limitations of this approach are:
\begin{itemize}
\item 
   Even the LO Monte-Carlo integration of~(\ref{eq:fnmain})
   is a trade-off between speed  
   and precision. With finite statistical errors,
   however, theory predictions are not ideally smooth 
   functions of the fit parameters.
   This contributes to numerical noise in the $\chi^2$
   calculations~\cite{Pumplin:2000vx}
   distorting the  $\chi^2$ contour during the  
   PDF error analysis, especially for fit parameters 
   with small errors.

\item
   The procedure can only be used for observables for 
   which LO calculations  are fast. 
   Currently, this prevents the global PDF analyses from  
   using Tevatron dijet data and DIS jet data. 
\end{itemize}
In a time when phenomenology is aiming towards 
NNLO precision~\cite{Alekhin:2005gq,Martin:2004ir},
the $k$-factor approximation is clearly not satisfying concerning both
its limitation  in precision and its restrictions concerning data sets.

% --------------------------------------------------------------
% --------------------------------------------------------------
% --------------------------------------------------------------
\subsection{The fastNLO Solution}

\begin{figure}[t]
\centerline{
   \psfig{figure=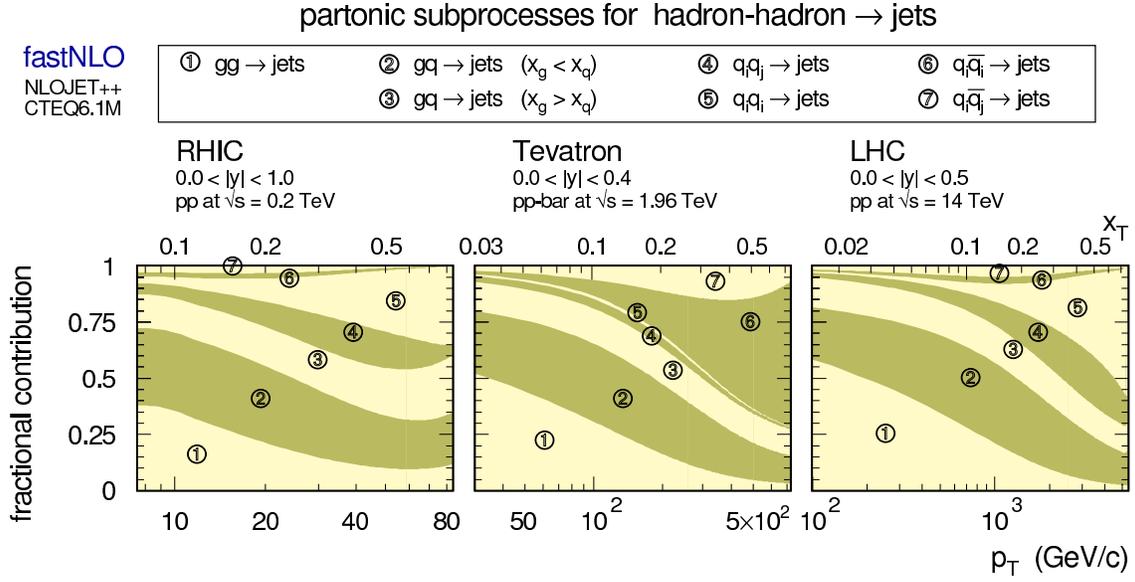,width=15cm}
}
  \caption{Contributions of different partonic subprocesses to 
   the inclusive jet cross section at 
   RHIC (left), the Tevatron (middle) and the LHC (right)
   as a function of $p_T$ and $x_T = 2 p_T/\sqrt{s}$.
   The subprocess $gq \rightarrow {\rm jets}$ has been
   separated into the contributions (2) and (3) where either the 
   quark- or the gluon momentum fraction is larger.
  \label{fig:fnsubprocpp}}
\end{figure}

\begin{figure}[t]
\centerline{
   \psfig{figure=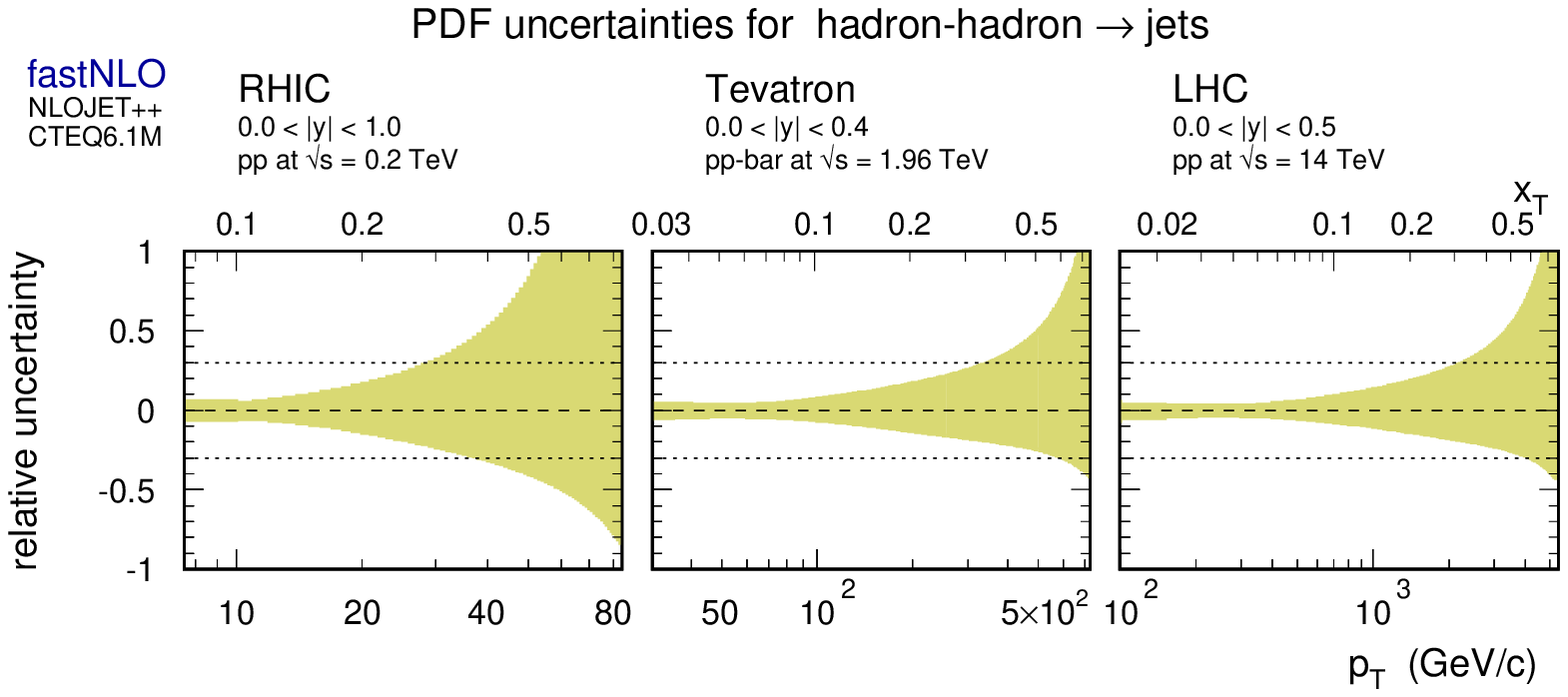,width=15cm}
}
  \caption{Comparison of PDF uncertainties for 
   the inclusive jet cross section at 
   RHIC (left), the Tevatron (middle) and the LHC (right).
   The uncertainty band is obtained for the CTEQ6.1M 
   parton density functions and the results are shown
   as a function of $p_T$ and $x_T = 2 p_T/\sqrt{s}$.
  \label{fig:fnpdfuncpp}}
\end{figure}

A better solution is implemented in the fastNLO project.
The basic idea is to transform the convolution 
in~(\ref{eq:fnmain}) into the factorized expression~(\ref{eq:fnfinal}).
Many proposals for this have been made in the past, originally
related to solving the DGLAP parton evolution equations~\cite{Pascaud:1994vx}
and later to computing of jet cross 
sections~\cite{Lobo:1996,Graudenz:1995sk,Kosower:1997vj,Wobisch:2000dk,Carli:2005ji}.
The fastNLO method is an extension of the 
concepts developed for DIS jet production~\cite{Lobo:1996,Wobisch:2000dk}
which have been applied at HERA
to determine the gluon density in the proton from DIS jet data~\cite{Adloff:2000tq}.
Starting from~(\ref{eq:fnmain}) for  the following discussion the 
renormalization scale is set equal to the factorization scale 
($\mu_{r,f}=\mu$).
The extension to $\mu_r \ne \mu_f$ is, however, trivial.
The $x$ dependence of the PDFs and the 
scale dependence of $\alpha_s^n$ and the PDFs can be approximated 
using an interpolation between sets of fixed values $x^{(k)}$ 
and $\mu^{(m)}$ 
($k=1, \cdots, k_{\rm max}\,;\,  m =1, \cdots, m_{\rm max}$)
\begin{eqnarray}
  &  \alpha^n_s(\mu)&  \cdot \; F_i(x_a,x_b,\mu) \; \simeq 
      \hskip28mm 
{[{\scriptstyle \mbox{``='' is true for 
$k_{\rm max}, l_{\rm max}, m_{\rm max}\rightarrow \infty $} }]}
\nonumber \\
& & 
\sum_{k,l,m}  \alpha^n_s(\mu^{(m)}) \cdot F_i(x_a^{(k)}, x_b^{(l)}, \mu^{(m)}) 
\, \cdot \,  e^{(k)}(x_a) \cdot  e^{(l)}(x_b) \cdot b^{(m)}(\mu)   
\end{eqnarray}
where $e^{(k,l)}(x)$ and $b^{(m)}(\mu)$ are interpolation functions
for the $x$ and the $\mu$ dependence, respectively.
All information of the perturbatively calculable piece
(including phase space restrictions, jet definition, etc.\
but excluding $\alpha_s$ and the PDFs)
is fully contained in the quantity
\begin{equation}
\tilde{\sigma}_{n,i,k,l,m}(\mu) = 
 c_{n,i}(x_a, x_b, \mu) \otimes 
\left[ e^{(k)}(x_a) \cdot e^{(l)}(x_b)  \cdot b^{(m)}(\mu) \right] \, .
\label{eq:sigmatilde}
\end{equation}
In the final prediction for the cross section
the convolution in~(\ref{eq:fnmain}) is then reduced
to a simple product
\begin{equation}
\sigma(\mu) \, \simeq \sum_{n,i,k,l,m} 
\tilde{\sigma}_{n,i,k,l,m}(\mu)  \, \cdot \,
 \alpha^n_s(\mu^{(m)}) \cdot
 F_i(x_a^{(k)}, x_b^{(l)}, \mu^{(m)}) \, .
\label{eq:fnfinal}
\end{equation}
The time-consuming step involving the calculation of the universal
(PDF and $\alpha_s$ independent) $\tilde\sigma$
is therefore factorized and needs to be done only once.
Any further calculation of the pQCD prediction
for arbitrary PDFs and $\alpha_s$ values can later
be done very fast by computing the simple sum of products
in~(\ref{eq:fnfinal}).
While the extension of the method from one 
initial-state hadron~\cite{Wobisch:2000dk}
to two hadrons was conceptually trivial, the case of two hadrons
requires additional efforts to improve the efficiency
and precision of the interpolation.
Both, the efficiency and the precision, are directly related to the 
choices of the points 
$x^{(k,l)}$, $\mu^{(m)}$ and the 
interpolation functions $e(x)$, $b(\mu)$.
The implementation in 
fastNLO achieves a precision of better than $0.1\%$ 
for $k_{\rm max},l_{\rm max} =10$ and $m_{\rm max}\le 4$.
Computation times for cross sections in fastNLO are roughly 
40-200\,$\mu$s per order $\alpha_s$ (depending on 
$m_{\rm max}$).
Further details are given in Ref~\cite{fastnlo}.

The $\tilde{\sigma}$ in~(\ref{eq:sigmatilde}) are computed using 
{\tt NLOJET++}~\cite{Nagy:2003tz,Nagy:2001fj}.
A unique feature in fastNLO is the inclusion of the $O(\alpha_s^4)$
threshold correction terms to the 
inclusive jet cross section~\cite{Kidonakis:2000gi},
a first step towards a full NNLO calculation.

% --------------------------------------------------------------
% --------------------------------------------------------------
% --------------------------------------------------------------
\section{Results}

\begin{figure}[!h]
\centerline{
   \psfig{figure=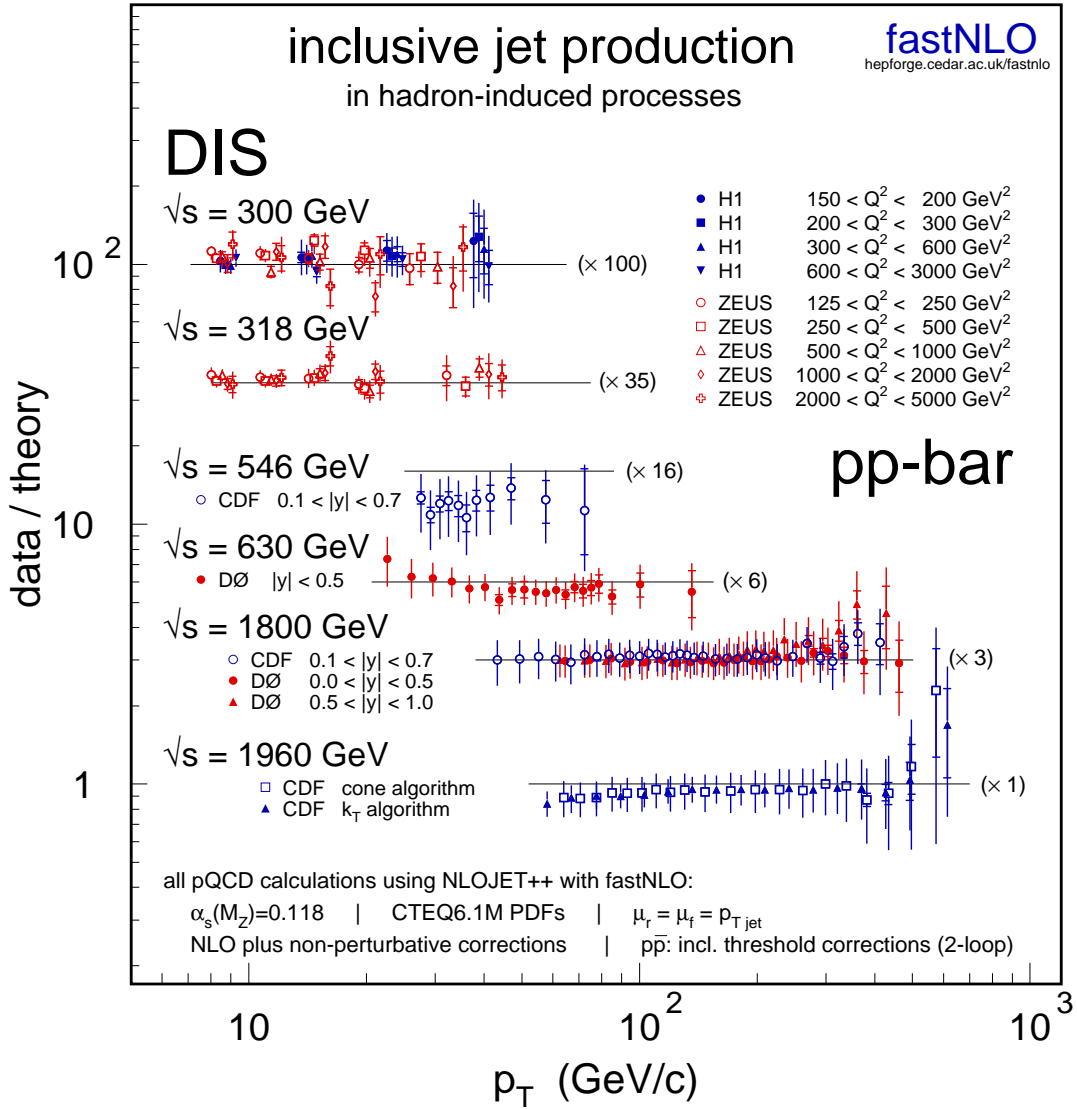,width=14.3cm}
}
  \caption{An overview of data over theory ratios for 
  inclusive jet cross sections, measured 
  in different processes at different center-of-mass energies.
  The data are compared to calculations obtained by fastNLO
  in NLO precision (for DIS data) and including 
  ${\cal O}(\alpha_s^4)$ threshold  corrections (for $p\bar{p}$ data).
  The inner error bars represent the statistical errors and the
  outer error bars correspond to the quadratic sum of all 
  experimental uncertainties.
  In all cases the perturbative predictions have been 
  corrected for non-perturbative effects.
  \label{fig:fnresults}}
\end{figure}

\begin{figure}[!h]
\centerline{
   \psfig{figure=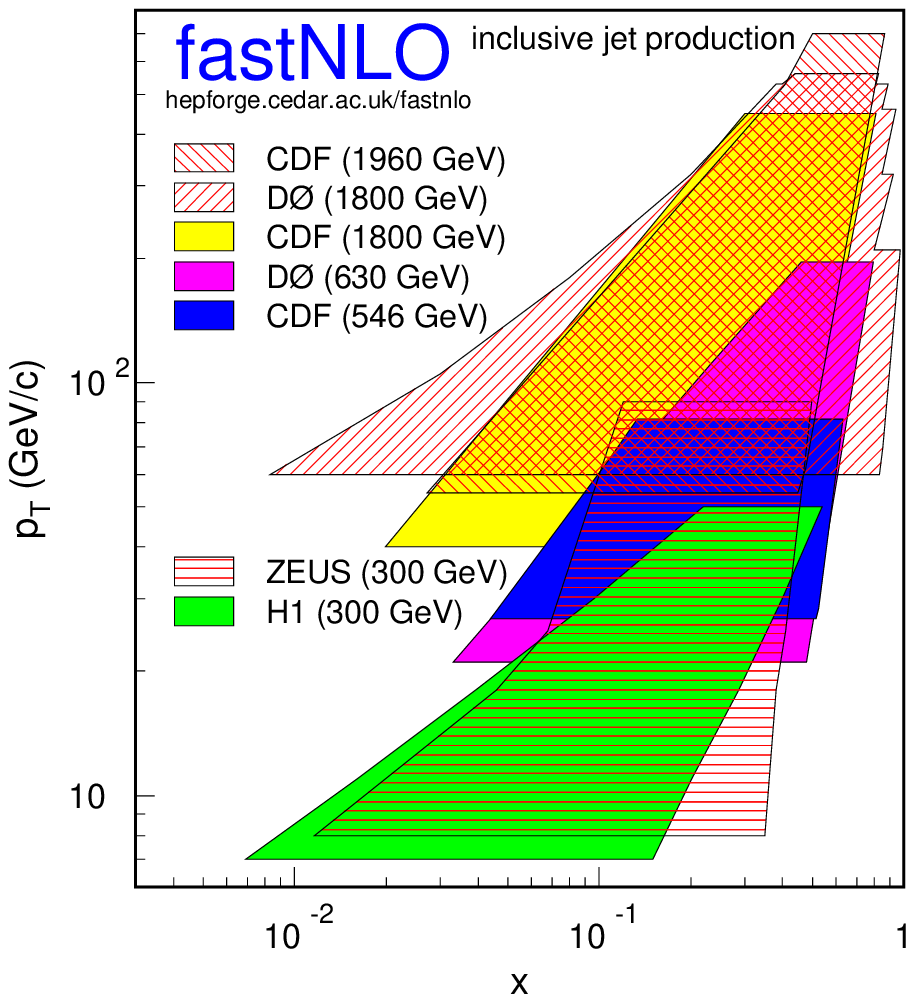,width=11cm}
}
  \caption{The phase space in $x$ and $p_T$
      covered by the data sets shown in the previous figure.
  \label{fig:fnresults2}}
\end{figure}

Calculations by fastNLO
are available at {\tt http://hepforge.cedar.ac.uk/fastnlo}
for a large set of (published, ongoing, or planned) 
jet cross section measurements at 
HERA, RHIC, the Tevatron, and the LHC
(either online or as computer code for inclusion in PDF fits).
Some fastNLO results for the inclusive jet cross section 
in different reactions are shown in this section.
The contributions from different partonic subprocesses
to the central inclusive jet cross section
are compared in Fig.~\ref{fig:fnsubprocpp} for different
colliders: 
For $pp$ collisions at RHIC and the LHC, 
and for $p\bar{p}$ scattering at Tevatron Run II energies.
It is seen that the quark-induced subprocesses are dominated
by the valence quarks:
In proton-proton collisions (RHIC, LHC)
the quark-quark subprocesses (4,5) give much larger 
contributions than the quark-antiquark subprocesses (6,7)
while exactly the opposite is true for proton-antiproton collisions
at the Tevatron.
The contribution from gluon-induced subprocesses is 
significant at all colliders over the whole $p_T$ ranges.
It is interesting to note that at fixed $x_T = 2 p_T/\sqrt{s}$
the gluon contributions are largest at RHIC.
Here, the jet cross section at 
$x_T = 0.5$ still receives $55\%$
contributions from gluon-induced subprocesses,
as compared to only $35\%$ at the Tevatron or $38\%$ at the LHC.
As shown in Fig.~\ref{fig:fnpdfuncpp}, this results in much larger
PDF uncertainties for the high $x_T$ inclusive jet cross section 
at RHIC, as compared to the Tevatron and the LHC for which
PDF uncertainties are roughly 
of the same size (at the same $x_T$).
This indicates that the PDF sensitivity at the same $x_T$
is about the same at the Tevatron and at the LHC, 
while it is much higher at RHIC.

An overview over published measurements of the inclusive
jet cross section in different reactions and at different
center-of-mass energies is given in Fig.~\ref{fig:fnresults}.
The results are shown as ratios of data over theory.
The theory calculations include the best available 
perturbative predictions (NLO for DIS data and NLO + 
${\cal O}(\alpha_s^4)$ threshold corrections for $p\bar{p}$ data)
which have been corrected for non-perturbative effects.
Over the whole phase space of $8 < p_T < 700$\,GeV
jet data in DIS and $p\bar{p}$ collisions are well-described
by the theory predictions using 
CTEQ6.1M PDFs~\cite{Pumplin:2002vw}.
The phase space in $x$ and $p_T$ covered 
by these measurements is shown in Fig.~\ref{fig:fnresults2},
demonstrating what can be gained by using fastNLO 
to include these data sets in future PDF fits.
A first study using fastNLO on the future potential
of LHC jet data has been published in Ref.~\cite{cmsptdrv2}.

\end{document}